\documentclass[11pt, a4paper]{article}
\usepackage[T1]{fontenc}
\usepackage{mathptmx}
\usepackage[a4paper, margin=2.2cm]{geometry}
\usepackage{amsmath}
\usepackage{booktabs}
\usepackage{tabularx}
\usepackage{textcomp}
\usepackage{hyperref}

\hypersetup{
    colorlinks=true,
    linkcolor=blue,
    citecolor=blue,
    urlcolor=blue
}

\title{\textbf{Distribution and density of Brachyuran crabs in Shatt Al-Basra Canal, Basrah, Iraq}}

\author{
    Anfas N. Okash\textsuperscript{1,*}, 
    Fadya M. Saleem\textsuperscript{1}, 
    Sudad Asaad Alkinani\textsuperscript{1}, 
    Zahraa M. Abdulhameed\textsuperscript{2} \\[6pt]
    \small \textsuperscript{1}Department of Ecology, College of Science, University of Basrah, Basrah, Iraq. \\
    \small \textsuperscript{2}Department of Architecture, College of Engineering, University of Basrah, Basrah, Iraq. \\
    \small \textsuperscript{*}Corresponding author: \texttt{anfas.okash@uobasrah.edu.iq}
}

\date{}

\begin{document}

\maketitle

\begin{abstract}
Crab samples were collected monthly from September 2023 to June 2024 in the Shatt Al-Basra Canal, Basrah, Iraq. During sampling, the air temperature ranged from 8 to $38^\circ\text{C}$, biological oxygen demand (BOD) from 3.5 to $7.3\text{ mg/L}$, and total organic carbon (TOC) from 0.25 to 0.54. Based on the results, the highest density recorded for \textit{Nasima dotilliformis} in December was $6\text{ ind/m}^2$, for \textit{Leptochryseus kuwaitensis} in February, and for \textit{Ilyoplax stevensi} in December and January, reaching $4\text{ ind/m}^2$. \textit{Nasima dotilliformis} was found throughout the whole study period in all months, but the two other species were absent in some months, with \textit{L. kuwaitensis} having a lower density. Temperature had the most significant impact on the presence and density of the crab community and was a fundamental factor in the dryness of the riverbank soil. The results also showed a significant inverse correlation between temperature and BOD and an insignificant direct correlation between temperature and TOC.

\vspace{6pt}
\end{abstract}

\section{Introduction}
Crabs have a well-developed carapace, usually wider than the length and depth of the crabs' body, with the abdomen folded underneath as a segmented flap and the first pair of clawed periopods \cite{Thomson1951}. The predatory and mutualistic relationships between crabs and other marine animals, reproductive biology, population dynamics, and their importance are fascinating to marine ecologists. Various aspects of crab behavior, such as burrowing, sound production, and foraging, are of interest to animal behaviorists. The physiological adaptations of their osmotic balance, respiration and ventilation, hormonal control of molting, autotomy, and regeneration of lost limbs, and highly organized nervous systems are exciting for physiologists \cite{McLay1988}. Many benthic crabs pass through a planktonic molting stage, which is determined by the number of molts, the duration of molting, and the increase in body size at molting. They then metamorphose into megalopa, during which the transition from plankton to benthic life occurs \cite{Anger2006}.

Recently, great attention has been paid to studying the brachyuran fauna of Iraq and their identification and distribution \cite{Naser2009, Naser2013, Yasser2019}, and many studies conducted along the coastal region of the Persian Gulf \cite{Jones1986, AlKhayat1996, Apel1996, AlGhais1996, Hornby1997}. Brachyuran crab species are common inhabitants of rocky shores \cite{Ng2008}. Yasser and Naser \cite{Yasser2021} listed 30 species on the Iraqi marine coasts that belong to the Brachyuran crab, most of which belong to the family Pilumnidae, in four genera and five species, followed by Leucosiidae, Camptandriidae, Macrophthalmidae, Sesarmidae, Epialtidae, and Varunidae, respectively. This study aimed to investigate the distribution and density of Brachyuran crabs in Shatt Al-Basra Canal, Basrah, Iraq.

\section{Materials and Methods}
Samples were collected monthly from September 2023 to June 2024 from the banks of the Shatt Al-Basra Canal, beyond the canal's regulator ($30^\circ 24' 7''\text{ N}, 47^\circ 46' 49''\text{ E}$). The samples were collected using a plastic quadrate with a $1\text{ m}$ length, which was thrown 16 times to cover the entire study stations. Each square containing the crab nest holes was dug to a depth of $15\text{ cm}$ to collect the specimens. After being identified, the crabs were returned to their habitats, and some samples were fixed and preserved in 10\% buffered formalin. Soil sampling was also done to estimate total organic carbon and texture and to correlate crab densities with soil structure. In addition, air temperature, biological oxygen demand (BOD), and total organic carbon (TOC) of the studied stations in the Shatt Al-Basra Canal were measured during sampling.

\section{Results}
\subsection{Study Area and Environmental Parameters}
The Shatt Al-Basra River is a sewage channel for all regions of Basra Governorate. The sampling area was characterized by high levels of pollution, which led to a lack of biodiversity in the area, except for a few individuals of polychaetes and mud skippers. However, due to the availability of organic materials, many insects are found, followed by local and migratory birds.

The highest temperature during the sampling period was $38^\circ\text{C}$ in June 2024, while the lowest temperature was $8^\circ\text{C}$ in January. The highest BOD value was recorded in December 2023 and March 2024, when it reached $7.3\text{ mg/L}$, while the lowest value was recorded in October, when it reached $3.5\text{ mg/L}$. The highest TOC value was recorded in May, when it reached 0.54 (54\%), while the lowest was in February, when it reached 0.25 (25\%) (Table~\ref{tab:env}).

\begin{table}[ht]
\centering
\caption{Values of air temperature ($^\circ\text{C}$), BOD ($\text{mg/L}$), and TOC during the study period.}
\label{tab:env}
\begin{tabular}{lcccccccccc}
\toprule
\textbf{Parameter} & \textbf{Sep.} & \textbf{Oct.} & \textbf{Nov.} & \textbf{Dec.} & \textbf{Jan.} & \textbf{Feb.} & \textbf{Mar.} & \textbf{Apr.} & \textbf{May} & \textbf{Jun.} \\
\midrule
Air temp. ($^\circ\text{C}$) & 30 & 28 & 24 & 11 & 8 & 13 & 21 & 33 & 34 & 38 \\
BOD ($\text{mg/L}$) & 3.6 & 3.5 & 4.6 & 7.3 & 7.0 & 6.4 & 7.3 & 6.0 & 3.9 & 3.6 \\
TOC & 0.33 & 0.29 & 0.33 & 0.45 & 0.34 & 0.25 & 0.49 & 0.31 & 0.54 & 0.53 \\
\bottomrule
\end{tabular}
\end{table}

\subsection{Crab Distribution and Density}
Three Brachyuran crabs were collected during the study period from Shatt Al-Basra Canal, viz. \textit{Nasima dotilliformis}, \textit{Leptochryseus kuwaitensis}, and \textit{Ilyoplax stevensi}. The density of \textit{N. dotilliformis} ranged between 2 and $6\text{ ind/m}^2$, \textit{L. kuwaitensis} $0\text{--}4\text{ ind/m}^2$, and \textit{I. stevensi} $0\text{--}4\text{ ind/m}^2$ in the sampling stations (Table~\ref{tab:density}). The highest presence ratio was for \textit{N. dotilliformis} in June, where it reached 100\%, while the lowest presence ratio was for \textit{L. kuwaitensis} in January, where it reached 10\%. However, the ratio was 0\% for \textit{L. kuwaitensis} in other months.

\begin{table}[ht]
\centering
\caption{Monthly density ($\text{ind/m}^2$) of the three Brachyuran crab species during the study period.}
\label{tab:density}
\begin{tabular}{lcccccccccc}
\toprule
\textbf{Species} & \textbf{Sep.} & \textbf{Oct.} & \textbf{Nov.} & \textbf{Dec.} & \textbf{Jan.} & \textbf{Feb.} & \textbf{Mar.} & \textbf{Apr.} & \textbf{May} & \textbf{Jun.} \\
\midrule
\textit{Nasima dotilliformis} & 4 & 3 & 4 & 6 & 5 & 3 & 5 & 5 & 2 & 3 \\
\textit{Leptochryseus kuwaitensis} & 0 & 0 & 0 & 2 & 1 & 4 & 2 & 0 & 1 & 0 \\
\textit{Ilyoplax stevensi} & 3 & 2 & 2 & 4 & 4 & 2 & 3 & 2 & 0 & 0 \\
\midrule
\textbf{Total Density} & \textbf{7} & \textbf{5} & \textbf{6} & \textbf{12} & \textbf{10} & \textbf{9} & \textbf{10} & \textbf{7} & \textbf{3} & \textbf{3} \\
\bottomrule
\end{tabular}
\end{table}

The data indicate a negative relationship between the monthly total density of all crab species and the air temperature during the study period. In addition, the relationship between the BOD and the total monthly density of all crabs was significantly positive, whereas the relationship between organic carbon and crab density was insignificant.

\section{Discussion}
High temperatures in the summer characterize the climate of Iraq, which is temperate during the winter. There is also a tremendous thermal fluctuation between night and day. The Shatt Al-Basrah is an artificial channel for collecting sewage water from all areas of the Basrah Governorate. Also, most of its banks are a dumping ground for waste from nearby residential areas, construction crushers, and animal slaughterhouses, making it a harsh area for many organisms. Therefore, it is characterized by very poor biodiversity.

The current study finds \textit{L. kuwaitensis}, \textit{I. stevensi}, and \textit{N. dotilliformis} in the Shatt Al-Basrah. Azra et al. \cite{Azra2020} reported that Brachyura species are sensitive to temperature changes. Our results also revealed a significant inverse correlation between temperature and crabs' density for all collected species, agreeing with findings that crabs differ in their tolerance and response to temperature changes \cite{Azra2020}. Watson et al. \cite{Watson2018} found that \textit{Ocypode cordimanus} exhibits high-temperature tolerance by descending into burrows to depths up to $40\text{ cm}$.

Some studies have reported a positive correlation between BOD and temperature; however, our work found the opposite, an inverse correlation. The reason could be the dryness of the tidal zone, where the decomposition process occurs within a record time of the water receding, which negatively affects the decomposition of organic matter. Therefore, due to the moderate temperatures during the winter months and the high humidity, BOD rates increased during these months, leading to a positive correlation between the density of all crab species and BOD.

The results of this study were consistent with the findings of Chumsri et al. \cite{Chumsri2023}, who found that human disturbances affect the behavior of fiddler crab, \textit{Austruca annulipes}, which increases anti-predatory behaviors more than mating behaviors. Despite the correlation between crab density and BOD, the results did not show any correlation between crab density and TOC, consistent with the findings of Lam-Gordillo et al. \cite{LamGordillo2022}, who found that depth, salinity, temperature, and sand content were the most influential factors on crab density. The environment of the Shatt Al-Basra Canal suffers from many pollutants, as it is a sewage channel and dumping ground for its riverbanks, justifying a lack of biodiversity and the low density of existing organisms.

\end{document}